\documentclass[aps,prl,twocolumn,superscriptaddress,floatfix,showpacs,preprintnumbers,amsmath,amssymb,showkeys]{revtex4}

\usepackage{graphicx}
\usepackage{dcolumn}

\newcommand{\ameas}{\mbox{$A_{meas}$}} 
\newcommand{\aperp}{\mbox{$A_{\perp}$}} 
\newcommand{\avap}{\mbox{$A_{n}$}} 
\newcommand{\aelas}{\mbox{$A_{elas}$}} 
\newcommand{\abkg}{\mbox{$A_{bkg}$}} 

\begin{document}


\title{Transverse Beam Spin Asymmetries in Forward-Angle \\Elastic Electron-Proton Scattering}


\author{D.~S.~Armstrong}
\affiliation{Department of Physics, College of William and Mary, Williamsburg, Virginia 23187, USA}
\author{J.~Arvieux}
\affiliation{Institut de Physique Nucl\'eaire d'Orsay, CNRS/IN2P3, Universit\'e Paris Sud, Orsay, France}
\author{R.~Asaturyan}
\altaffiliation{Deceased.}
\affiliation{Yerevan Physics Institute, Yerevan 375036 Armenia}
\author{T.~Averett}
\affiliation{Department of Physics, College of William and Mary, Williamsburg, Virginia 23187, USA}
\author{S.~L.~Bailey}
\affiliation{Department of Physics, College of William and Mary, Williamsburg, Virginia 23187, USA}
\author{G.~Batigne}
\affiliation{LPSC, Universit\'e Joseph Fourier Grenoble 1, CNRS/IN2P3, INPG, Grenoble, France}
\author{D.~H.~Beck}
\affiliation{Loomis Laboratory of Physics, University of Illinois, Urbana, Illinois 61801, USA}
\author{E.~J.~Beise}
\affiliation{Physics Department, University of Maryland, College Park, Maryland 20472, USA}
\author{J.~Benesch}
\affiliation{Thomas Jefferson National Accelerator Facility, Newport News, Virginia, 23606, USA}
\author{L.~Bimbot}
\affiliation{Institut de Physique Nucl\'eaire d'Orsay, CNRS/IN2P3, Universit\'e Paris Sud, Orsay, France}
% ...

\collaboration{G0 Collaboration}
\noaffiliation

\date{\today}

\begin{abstract}
We have measured the beam-normal single-spin asymmetry in elastic scattering of transversely-polarized 3 GeV electrons from unpolarized protons at $Q^2 = 0.15, 0.25$ (GeV/$c$)$^2$.  The results are inconsistent with calculations solely using the elastic nucleon intermediate state, and generally agree with calculations with significant inelastic hadronic intermediate state contributions.  \avap{} provides a direct probe of the imaginary component of the 2$\gamma$ exchange amplitude, the complete description of which is important in the interpretation of data from precision electron-scattering experiments.
\end{abstract}



\maketitle


Elastic scattering of electrons from nucleons is usually treated in the single-photon exchange (Born) approximation.  Higher order processes, such as two-photon exchange, are generally treated as small radiative corrections.  However, interest in two-photon exchange was recently renewed when it was argued that contributions from the real part of this amplitude play a role in the discrepancy between the Rosenbluth separation and polarization transfer measurements of the ratio of the elastic form factors $G_E^p/G_M^p$~\cite{Qattan:2004ht,Punjabi:2005wq,Guichon:2003qm}.  In addition, although the  two-photon exchange contribution is small, it is comparable to the parity-violating elastic electron-nucleon scattering asymmetry~\cite{Afanasev:2005ex}, and recent parity-violation measurements have had to consider possible systematic corrections due to this effect.  A good understanding of two-photon exchange contributions can be extended to calculations of diagrams that appear in other processes, such as $\gamma Z$ and $W^+ W^-$ box diagrams, which are important corrections in precision electroweak experiments~\cite{Marciano:1983ss}. Thus, empirical verification of the theoretical framework for this effect is beneficial.

The two-photon exchange process involves the exchange of two virtual photons with an intermediate hadronic state that includes the ground state and all excited states, and can produce a single-spin asymmetry in electron scattering~\cite{DeRujula:1972te}.  The beam-normal single spin asymmetry, or transverse asymmetry \avap{}, is sensitive to the imaginary  
part of the two-photon exchange amplitude in the elastic scattering of transversely polarized electrons from unpolarized nucleons, and arises from the interference of the one-photon and two-photon exchange amplitudes~\cite{DeRujula:1972te}.  Time-reversal invariance forces \avap{} to vanish in the Born approximation, so it is of relative order $\alpha \approx \frac{1}{137}$.  Furthermore, $A_n$ must vanish in the chiral limit and so is suppressed by the ratio of the electron's rest mass to the beam energy, leading to an asymmetry on order of $10^{-5}$--$10^{-6}$ for $\simeq$ GeV electrons. Hence, measurement of \avap{} is challenging.  For a beam polarized normal to the scattering plane, the transverse asymmetry is defined as 
$A_n = \frac{\sigma_{\uparrow} - \sigma_{\downarrow}}{\sigma_{\uparrow} + \sigma_{\downarrow}}$,
where $\sigma_{\uparrow} (\sigma_{\downarrow})$ represents the cross section for the elastic scattering of electrons with spins parallel (anti-parallel) to the normal polarization vector defined by
$\hat{n} \equiv \frac{(\vec{k}_e \times \vec{k}')}{|\vec{k}_e \times \vec{k}'|}$, 
where $\vec{k}_e$ and $\vec{k}'$ are the three-momenta for the incident and scattered electron.  The measured asymmetry \ameas{} can be written as $A_n \vec{P}_e \cdot \hat{n}$.
Because of the term \mbox{$\vec{P}_e \cdot \hat{n}$}, \ameas{} is dependent on the azimuthal scattering angle $\phi$, which is manifested as a sinusoidal dependence in \ameas{} vs. $\phi$.

The transverse asymmetry due to two-photon exchange can be expressed using the formalism developed for the general amplitude for electron-nucleon elastic scattering~\cite{Goldberger:1957te}.  This parameterization uses six complex functions, $\tilde{G}_M(\nu,Q^2)$, $\tilde{G}_E(\nu,Q^2)$, and $\tilde{F}_i(\nu,Q^2)$, $i = 3,\ldots,6$, dependent on $\nu$, the energy transfer to the proton, and $Q^2$, the four-momentum transfer.  In the Born approximation, these functions reduce to the usual magnetic and electric form factors $G_M(Q^2)$, $G_E(Q^2)$, and to $\tilde{F}_i = 0$, so the $\tilde{F}_i$ and phases of $G_M$, $G_E$ must come from processes with the exchange of two or more photons.  \avap{} is proportional to the imaginary part of the combination of $\tilde{F}_3$, $\tilde{F}_4$, $\tilde{F}_5$~\cite{Pasquini:2004pv,Gorchtein:2004ac}.
Thus, $A_n$ is a function of $Q^2$ and the center-of-mass scattering angle $\theta_{CM}$, with the intermediate hadronic state information contained in the $\tilde{F}_i$.

 There have been several calculations of the transverse asymmetry for the present kinematics~\cite{Gorchtein:2004ac,Afanasev:2004pu,Gorchtein:2006mq,Pasquini:2004pv,Carlson:2007sp}, but the primary theoretical difficulty in calculations of the two-photon exchange amplitude is the large uncertainty in the contribution of the inelastic hadronic intermediate states.  As the calculations require both the proton elastic form factors (elastic contribution) and the excitation amplitudes to other intermediate states, e.g. $\pi N$ (inelastic contribution), experimental verification is important to test the framework of the calculations.  However, at present experimental information on \avap{} is scarce.

The first measurement of \avap{} was performed by the SAMPLE Collaboration, at laboratory scattering angles of $130^{\circ} \leq \theta_e \leq 170^{\circ}$ at $Q^2 = 0.1$ (GeV/$c$)$^2$ with a 200 MeV beam~\cite{Wells:2000rx}.
  The PVA4 Collaboration recently reported measurements at somewhat higher beam energies at $30^{\circ} \leq \theta_e \leq 40^{\circ}$ and $Q^2 = 0.106, 0.230$ (GeV/$c$)$^2$~\cite{Maas:2004pd}.  The results from both indicate that models including only the nucleon elastic state are insufficient.
Other preliminary data also suggest that the
elastic contribution alone is insufficient \cite{Capozza}.

The present measurement is at a higher beam energy (3 GeV) and forward angle,
where the $\pi N$ intermediate states are predicted to 
be a significant contribution to \avap{}~\cite{Pasquini:2004pv}.
Furthermore, this beam energy falls in a transition range between models.  At energies below the two-pion production threshold, the $\pi N$ intermediate state contribution can be calculated using pion electroproduction amplitudes based on experimental input.  Above that limit, the $\pi N$  contribution is not well known, and there could be additional contributions to $A_n$~\cite{Pasquini:2004pv}.  At very high energies and forward scattering angles (the diffractive limit), $A_n$ can be expressed simply in terms of the total photo-absorption cross section using the optical theorem~\cite{Afanasev:2004pu,Gorchtein:2005za}.  For the present intermediate energy, corrections to the diffractive limit result have been calculated~\cite{Afanasev:2004pu,Gorchtein:2006mq}.


We report a measurement of the beam-normal
spin asymmetry \avap{} from the elastic scattering of transversely polarized electrons from the proton at forward scattering angles using the 
apparatus for the G0 experiment~\cite{Armstrong:2005hs} 
in Hall C at Jefferson Lab.  The apparatus and the method of asymmetry extraction are described in Refs~\cite{Armstrong:2005hs,G0NIM,Covrig:2005vx,G0_electronics_NIM}; only differences between the running conditions and analysis of the transverse asymmetry data and the parity-violation data are reported here.

A 40 $\mu$A electron beam of energy \mbox{$3.031$ GeV} was generated with a strained GaAs polarized source~\cite{Sinclair:2007ez}.  The beam had a 32 ns 
pulse separation, to allow 
particle identification via time-of-flight (ToF).  
The beam helicity  
was held constant for $\frac{1}{30}$ s periods in sequences chosen pseudo-randomly.  Active feedback systems maintained 
helicity-correlated current and position changes below  
$\sim 0.3$ ppm and 10 nm, respectively.
A Wien filter   
precessed the spin of the longitudinally-polarized beam electrons in the 
accelerator plane.  The average longitudinal
beam polarization was measured with a M{\o}ller 
polarimeter~\cite{Hauger:1999iv} 
for a range of spin precessions, and these data were fit to
obtain the Wien setting for maximal transverse polarization in Hall C 
and to determine the magnitude of the polarization, $(74.3 \pm 1.3)$\%.  
The purely transverse nature of the beam was verified by a measurement of zero longitudinal polarization with the M{\o}ller polarimeter at the optimal Wien setting.

The polarized electrons were scattered from a \mbox{20 cm} long liquid hydrogen
target~\cite{Covrig:2005vx}.  Recoiling elastic protons were
focused by a superconducting magnet onto eight arrays of
segmented scintillation detectors  
arranged symmetrically in $\phi$ around the beam
axis~\cite{G0NIM}.  
Each array consisted of 16 scintillator pairs used in
coincidence to simultaneously measure the range of momentum transfers 
$0.12 \leq Q^2 \leq 1.0$  (GeV/$c$)$^2$.  The detectors were 
numbered in order of increasing values of $Q^2$. 
Detector events were sorted by ToF
using custom time-encoding electronics~\cite{G0_electronics_NIM}.

Thirty hours of data were taken in the transverse-beam
configuration.  The transverse asymmetry
is not expected to vary rapidly with $Q^2$ in
this kinematic region~\cite{Carlson:2007sp}, so in order to improve
statistical precision, the ToF
spectra from the first 
12 detectors in each array were summed into two spectra: 1--8 and 9--12.  These
correspond to ranges of $0.12 < Q^2 < 0.20$ and $0.20 < Q^2 < 0.32$ (GeV/$c$)$^2$, with 
central values of $Q^2 = 0.15$ (GeV/$c$)$^2$ and $\theta_{CM} = 20.2^{\circ}$, and $Q^2 = 0.25$ (GeV/$c$)$^2$ and
$\theta_{CM} = 25.9^{\circ}$, respectively.   
Data taken at higher $Q^2$ suffered from poor statistics and larger backgrounds, so results are not reported here.

\begin{figure}[t]
\includegraphics[width=5.5cm,angle=-90]{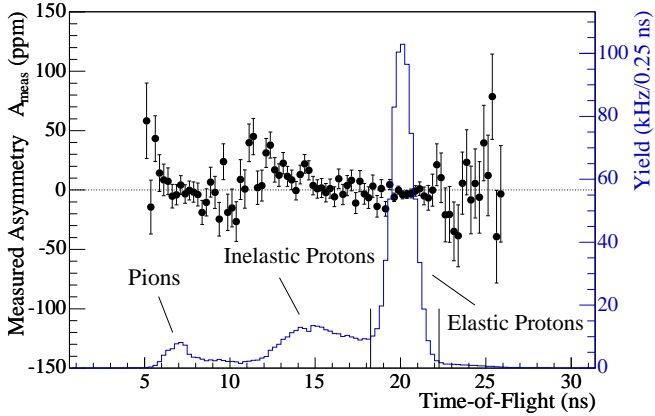} 
\caption{\label{fig:TOF} (Color online) Measured raw asymmetry, $\ameas{}$, (data points) and yield (histogram) as a function of the ToF for summed detectors 1--8 in the array located at $\langle \phi \rangle = 135^{\circ}$.  Error bars are statistical.}
\end{figure}

The rates for the individual detectors
were corrected for electronics deadtimes of 10\%--15\%,  
giving an uncertainty of $0.05$ ppm in the asymmetries.  
False asymmetries due to residual helicity-correlated beam current,
position, angle, and energy variations were calculated by linear 
regression to be $< \!\!0.12$ ppm. No correction was applied, and 
a systematic uncertainty of 0.12 ppm was adopted.  A $\phi$-independent 
correction of, on average, $+2.29 \pm 0.59$ ppm was made to account for
the large charge asymmetry ($\sim$570 ppm) of a small fraction
($\sim$$10^{-3}$) of the beam current with a 2 ns structure,
``leakage beam'' from beam intended for the other operating halls~\cite{G0NIM}.  Uncertainties of 0.002 ppm and 0.021 ppm for $Q^2 = 0.15$, $0.25$ (GeV/$c$)$^2$, respectively, arose due to the upper limit on 
the residual longitudinal polarization of the beam
and the known parity-violating asymmetry~\cite{Armstrong:2005hs}.  No radiative corrections were applied~\cite{afanasev_rad}.

Figure~\ref{fig:TOF} shows a typical ToF spectrum;
background
extends on both sides of the elastic proton peak, 
consisting of quasi-elastic protons from the
aluminum target windows and inelastic protons from both the aluminum
and the hydrogen of the target.  The measured asymmetry therefore
consists of two components
\begin{equation}
\ameas{} = f \aelas{} + (1-f) \abkg{},
\end{equation}
where \aelas{} is the elastic asymmetry, \abkg{} is the background
asymmetry contribution, and $f$ is the signal-to-measured yield
fraction.  The ToF spectra were rebinned into several regions; fits over these ToF bin regions to both the yield and
asymmetry in the region of the elastic peak were used in the background
correction.  The yield was modeled with a Gaussian elastic peak and a
fourth-order polynomial background.  The asymmetry was modeled using a
linear background and a constant for the elastic component.  
In a few cases, the asymmetry of the 
background
was comparable in size to the measured asymmetry. 
As a check, the background asymmetries were also extracted using an
alternate two-step fitting procedure in which the asymmetries from the background regions on either side of the elastic peak were used to interpolate the asymmetry from background processes beneath the elastic peak.  
This background asymmetry was then used to correct the elastic asymmetry, giving consistent results with the previous method.
For a given azimuthal angle, the systematic
uncertainty on the background correction was estimated from the
dispersion of the extracted background asymmetries between the two
methods.

The elastic transverse asymmetries, \aperp{} (the elastic asymmetry \aelas{}
corrected for all effects), 
for the eight detector arrays in
each $Q^2$ bin are presented in Table \ref{tab:final_corrected_asyms}.   
The systematic errors
are given in Table
\ref{tab:syst_errors}; the error on the back\-ground correction clearly
 dominates and varies in $\phi$.
A conservative model-dependent systematic error due to finite $Q^2$ bin size is indicated in Table II.

\begin{table}
\begin{centering}  
\caption{\label{tab:final_corrected_asyms} Elastic transverse asymmetries and uncertainties vs.\ azimuthal scattering angle $\phi$.  Uncertainties are
statistical and individual systematic uncertainties, respectively (in Table
\ref{tab:syst_errors}); global systematic uncertainties are not included.}
\begin{ruledtabular}
\begin{tabular}{ccc}

	&\multicolumn{2}{c}{\rule[-1.5mm]{0mm}{4.5mm}Elastic Transverse Asymmetries $\aperp{}$  (ppm)}\\ 
\raisebox{1.7ex}[0pt]{$\phi$}  &\multicolumn{1}{c}{$Q^2 = 0.15$ (GeV/$c$)$^2$} 	&\multicolumn{1}{c}{$Q^2 = 0.25$ (GeV/$c$)$^2$}\\ \hline 
0$^{\circ}$	&$-0.38 \pm 1.94\pm0.62$	&$\;\;\:1.39  \pm 6.22\pm2.36$\rule[0mm]{0mm}{4mm} \\
45$^{\circ}$	&$-1.15 \pm 1.84\pm0.49$	&$-1.09 \pm 3.35\pm1.09$ \\ 
90$^{\circ}$	&$-4.57 \pm 2.00\pm0.67$	&$-8.70 \pm 4.78\pm3.14$\\
135$^{\circ}$	&$-4.39 \pm 2.04\pm0.76$	&$-1.67 \pm 3.18\pm0.87$ \\
180$^{\circ}$	&$-3.30 \pm 1.88\pm0.67$	&$-6.45 \pm 5.70\pm3.86$ \\
225$^{\circ}$	&$\;\;\:2.74 \pm 1.88\pm0.39$	&$\;\;\:5.68  \pm 3.33\pm0.83$ \\
270$^{\circ}$	&$\;\;\:2.25 \pm 2.00\pm0.71$	&$\;\;\:9.74  \pm 4.58\pm2.75$ \\
315$^{\circ}$	&$\;\;\:4.47 \pm 1.98\pm0.42$	&$\;\;\:1.13  \pm 3.06\pm0.81$\\
\end{tabular}
\end{ruledtabular}
\end{centering}  
\end{table}

\begin{table}[b] 
\caption{\label{tab:syst_errors} Systematic uncertainties in the asymmetries. The first three are point-to-point; the last four are global.}
\begin{ruledtabular}                         
\begin{tabular}{ccc}
	&\multicolumn{2}{c}{Uncertainty}\\ 
\raisebox{1.7ex}[0pt]{Source\rule[-2mm]{0mm}{6mm} } 	&$Q^2 = 0.15$ GeV$^2$ 	&$Q^2 = 0.25$ GeV$^2$\\ \hline 
Deadtime \rule[0mm]{0mm}{4mm}		&0.05 ppm  	&0.05 ppm \\   
False asymmetries			&$0.12$ ppm 	&$0.12$ ppm \\
Background correction		&0.37 - 0.74 ppm 	& 0.80 - 3.86 ppm \\[1.2ex]
Leakage beam				&$0.55$ ppm  	&$0.63$ ppm	\\
Beam polarization			&1.8\% 		&1.8\% \\
Longitudinal polarization		&0.002 ppm  	&0.021 ppm\\
Finite $Q^2$ binning			&0.03 ppm  	&0.03 ppm\\
\end{tabular}
\end{ruledtabular}
\end{table}

The detector arrays are evenly-spaced in azimuthal
angle $\phi$ around the beamline, so the asymmetries should follow a
sinusoidal dependence in $\phi$, {\it viz.}
\begin{equation}
\label{eq:sine_dep}
\aperp{} = |\avap{}|\sin(\phi+\phi_0),
\end{equation}
where the amplitude $|\avap{}|$ is the magnitude of the transverse
 asymmetry, and the phase $\phi_0$ depends on the direction of the transverse beam polarization.   The electron polarization for the positive ($+$) helicity is in the beam left direction.  The azimuthal angle $\phi$ is defined to be $0^{\circ}$ on beam left and to increase clockwise as viewed along the direction of beam momentum.  
%
%
%
\begin{figure}[t]
\includegraphics[width=8.4cm,angle=0]{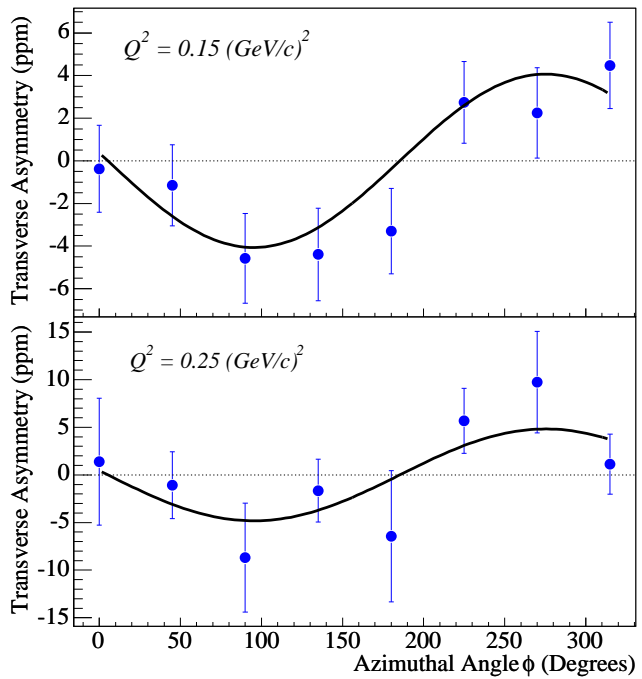}
\caption{\label{fig:sine_fits} (Color online) Measured
asymmetry as a function of the azimuthal
scattering angle $\phi$ for $Q^2 = 0.15$ (upper plot) and $0.25$ (GeV/$c$)$^2$ 
(lower plot). The curves are the best fit
to Eq.~\ref{eq:sine_dep}. Error bars are the statistical 
and individual systematic errors combined in quadrature.}
\end{figure}
%
%
The data's sinusoidal dependence is displayed in Fig.~\ref{fig:sine_fits}, 
along with the best
fit to Eq.~\ref{eq:sine_dep}.  
The fits were constrained to
$\phi_0$ as calculated from the electron spin precession 
in the 3T solenoid used in the M{\o}ller polarimeter (5.3$^{\circ}$) 
and do not allow a $\phi$-independent offset; however, relaxing
these constraints has negligible impact on the extracted $\avap{}$.  
The small effect of finite bin size in $\phi$ was included.

\begin{figure}[t]
\includegraphics[width=5.6cm,angle=-90]{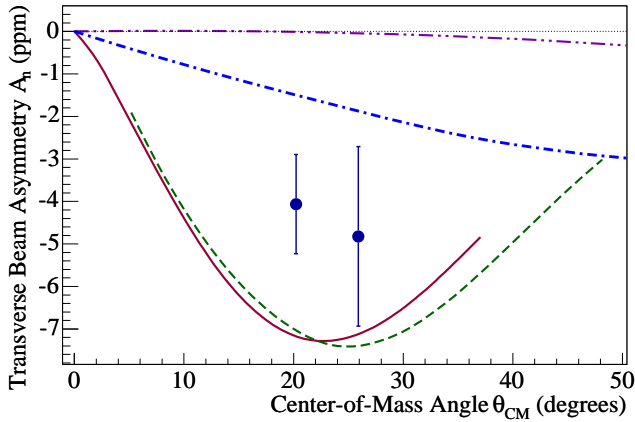}
\caption{\label{fig:An_thetaCM} (Color online) Results for $\avap{}$ as a function of center-of-mass scattering angle, along with calculations from Refs.~\cite{Afanasev:2004pu,Gorchtein:2006mq,Pasquini:2004pv} (see text for explanation). 
}
\end{figure}

Including all corrections, we obtain $A_n = -4.06 \pm 0.99_{\rm{stat}} \pm 0.63_{\rm{syst}}$ ppm for $Q^2 = 0.15$ (GeV/$c$)$^2$ and $-4.82 \pm 1.87_{\rm{stat}} \pm 0.98_{\rm{syst}}$ ppm for $Q^2 = 0.25$ (GeV/$c$)$^2$ from the sinusoidal fits.
Figure~\ref{fig:An_thetaCM} compares these results to available calculations~\cite{Afanasev:2004pu,Gorchtein:2006mq,Pasquini:2004pv} appropriate to these kinematics.  The dash-double-dotted line~\cite{Pasquini:2004pv} is a calculation of the two-photon exchange contribution solely from the nucleon intermediate state (elastic contribution); the dash-dotted line~\cite{Pasquini:2004pv} represents the intermediate hadronic state for which the elastic contribution  
has been combined with inelastic contributions from excitation amplitudes to $\pi N$-intermediate states.  The solid line~\cite{Afanasev:2004pu} and the dashed line~\cite{Gorchtein:2006mq} represent calculations using the optical theorem and parameterizations for the measured total photo-production cross sections on the proton.  
Clearly, the data show that the contribution of the inelastic hadronic intermediate states to the two-photon exchange amplitude is significant.  This conclusion is consistent with those reported by SAMPLE~\cite{Wells:2000rx} and PVA4~\cite{Maas:2004pd}; however, as the kinematics are different, the data points cannot be compared directly.  The G0 experiment and other experiments~\cite{Capozza} have recently obtained transverse beam spin asymmetry data at various angles on hydrogen, deuterium, and helium targets at additional $Q^2$ values, which will provide a further exploration of the imaginary part of the two-photon exchange amplitude.

The data reported here, along with other measurements, 
 provide a valuable test of the theoretical framework of the two-photon contribution to the cross section through a comparison of the 
 measured imaginary part of the two-photon exchange contribution to calculations of the real part, as the two are related through dispersion relations~\cite{Carlson:2007sp}.  These data underline the major role played by hadronic intermediate states in this process.  
Two-photon exchange and other box diagrams are important in the interpretation of high-precision parity-violating electron-scattering experiments and in the radiative corrections for other lepton scattering experiments, making an understanding of these contributions important.

This work is supported in part by
CNRS (France), DOE (USA), NSERC (Canada), and NSF (USA).
We thank A.~V.~Afanasev, M.~Gorchtein, B.~Pasquini, and M. Vanderhaeghen for their calculations and useful discussions.


\begin{thebibliography}{24}
\expandafter\ifx\csname natexlab\endcsname\relax\def\natexlab#1{#1}\fi
\expandafter\ifx\csname bibnamefont\endcsname\relax
  \def\bibnamefont#1{#1}\fi
\expandafter\ifx\csname bibfnamefont\endcsname\relax
  \def\bibfnamefont#1{#1}\fi
\expandafter\ifx\csname citenamefont\endcsname\relax
  \def\citenamefont#1{#1}\fi
\expandafter\ifx\csname url\endcsname\relax
  \def\url#1{\texttt{#1}}\fi
\expandafter\ifx\csname urlprefix\endcsname\relax\def\urlprefix{URL }\fi
\providecommand{\bibinfo}[2]{#2}
\providecommand{\eprint}[2][]{\url{#2}}

\bibitem[{\citenamefont{Qattan et~al.}(2005)}]{Qattan:2004ht}
\bibinfo{author}{\bibfnamefont{I.~A.} \bibnamefont{Qattan}}
  \bibnamefont{et~al.}, \bibinfo{journal}{Phys. Rev. Lett.}
  \textbf{\bibinfo{volume}{94}}, \bibinfo{pages}{142301}
  (\bibinfo{year}{2005}).

\bibitem[{\citenamefont{Punjabi et~al.}(2005)}]{Punjabi:2005wq}
\bibinfo{author}{\bibfnamefont{V.}~\bibnamefont{Punjabi}} \bibnamefont{et~al.},
  \bibinfo{journal}{Phys. Rev.} \textbf{\bibinfo{volume}{C71}},
  \bibinfo{pages}{055202} (\bibinfo{year}{2005}), \eprint{[Erratum-ibid.\ C
  {\bf 71}, 069902 (2005)]}.

\bibitem[{\citenamefont{Guichon and Vanderhaeghen}(2003)}]{Guichon:2003qm}
\bibinfo{author}{\bibfnamefont{P.~A.~M.} \bibnamefont{Guichon}}
  \bibnamefont{and}
  \bibinfo{author}{\bibfnamefont{M.}~\bibnamefont{Vanderhaeghen}},
  \bibinfo{journal}{Phys. Rev. Lett.} \textbf{\bibinfo{volume}{91}},
  \bibinfo{pages}{142303} (\bibinfo{year}{2003}).

\bibitem[{\citenamefont{Afanasev and Carlson}(2005)}]{Afanasev:2005ex}
\bibinfo{author}{\bibfnamefont{A.~V.} \bibnamefont{Afanasev}} \bibnamefont{and}
  \bibinfo{author}{\bibfnamefont{C.~E.} \bibnamefont{Carlson}},
  \bibinfo{journal}{Phys. Rev. Lett.} \textbf{\bibinfo{volume}{94}},
  \bibinfo{pages}{212301} (\bibinfo{year}{2005}).

\bibitem[{\citenamefont{Marciano and Sirlin}(1984 [Erratum-ibid. D 31:213,
  1985])}]{Marciano:1983ss}
\bibinfo{author}{\bibfnamefont{W.~J.} \bibnamefont{Marciano}} \bibnamefont{and}
  \bibinfo{author}{\bibfnamefont{A.}~\bibnamefont{Sirlin}},
  \bibinfo{journal}{Phys. Rev.} \textbf{\bibinfo{volume}{D29}},
  \bibinfo{pages}{75} (\bibinfo{year}{1984}), \eprint{[Erratum-ibid.\ D
  {\bf 31}, 213 (1985)]}.

\bibitem[{\citenamefont{De~Rujula et~al.}(1971)\citenamefont{De~Rujula, Kaplan,
  and De~Rafael}}]{DeRujula:1972te}
\bibinfo{author}{\bibfnamefont{A.}~\bibnamefont{De~Rujula}},
  \bibinfo{author}{\bibfnamefont{J.~M.} \bibnamefont{Kaplan}},
  \bibnamefont{and}
  \bibinfo{author}{\bibfnamefont{E.}~\bibnamefont{De~Rafael}},
  \bibinfo{journal}{Nucl. Phys.} \textbf{\bibinfo{volume}{B35}},
  \bibinfo{pages}{365} (\bibinfo{year}{1971}).

\bibitem[{\citenamefont{Goldberger et~al.}(1957)\citenamefont{Goldberger,
  Nambu, and Oehme}}]{Goldberger:1957te}
\bibinfo{author}{\bibfnamefont{M.~L.} \bibnamefont{Goldberger}},
  \bibinfo{author}{\bibfnamefont{Y.}~\bibnamefont{Nambu}}, \bibnamefont{and}
  \bibinfo{author}{\bibfnamefont{R.}~\bibnamefont{Oehme}},
  \bibinfo{journal}{Annals of Physics} \textbf{\bibinfo{volume}{2}},
  \bibinfo{pages}{226} (\bibinfo{year}{1957}).

\bibitem[{\citenamefont{Pasquini and Vanderhaeghen}(2004)}]{Pasquini:2004pv}
\bibinfo{author}{\bibfnamefont{B.}~\bibnamefont{Pasquini}} \bibnamefont{and}
  \bibinfo{author}{\bibfnamefont{M.}~\bibnamefont{Vanderhaeghen}},
  \bibinfo{journal}{Phys. Rev.} \textbf{\bibinfo{volume}{C70}},
  \bibinfo{pages}{045206} (\bibinfo{year}{2004}).

\bibitem[{\citenamefont{Gorchtein et~al.}(2004)\citenamefont{Gorchtein,
  Guichon, and Vanderhaeghen}}]{Gorchtein:2004ac}
\bibinfo{author}{\bibfnamefont{M.}~\bibnamefont{Gorchtein}},
  \bibinfo{author}{\bibfnamefont{P.~A.~M.} \bibnamefont{Guichon}},
  \bibnamefont{and}
  \bibinfo{author}{\bibfnamefont{M.}~\bibnamefont{Vanderhaeghen}},
  \bibinfo{journal}{Nucl. Phys.} \textbf{\bibinfo{volume}{A741}},
  \bibinfo{pages}{234} (\bibinfo{year}{2004}).

\bibitem[{\citenamefont{Afanasev and Merenkov}(2004)}]{Afanasev:2004pu}
\bibinfo{author}{\bibfnamefont{A.~V.} \bibnamefont{Afanasev}} \bibnamefont{and}
  \bibinfo{author}{\bibfnamefont{N.~P.} \bibnamefont{Merenkov}},
  \bibinfo{journal}{Phys. Lett.} \textbf{\bibinfo{volume}{B599}},
  \bibinfo{pages}{48} (\bibinfo{year}{2004}), \eprint{Erratum in hep-ph/0407167
  v2}.

\bibitem[{\citenamefont{Gorchtein}(2007)}]{Gorchtein:2006mq}
\bibinfo{author}{\bibfnamefont{M.}~\bibnamefont{Gorchtein}},
  \bibinfo{journal}{Phys. Lett.} \textbf{\bibinfo{volume}{B644}},
  \bibinfo{pages}{322} (\bibinfo{year}{2007}).

\bibitem[{\citenamefont{Carlson and Vanderhaeghen}(2007)}]{Carlson:2007sp}
\bibinfo{author}{\bibfnamefont{C.~E.} \bibnamefont{Carlson}} \bibnamefont{and}
  \bibinfo{author}{\bibfnamefont{M.}~\bibnamefont{Vanderhaeghen}}
  (\bibinfo{year}{2007}), \eprint{hep-ph/0701272}.


\bibitem[{\citenamefont{Wells et~al.}(2001)}]{Wells:2000rx}
\bibinfo{author}{\bibfnamefont{S.~P.} \bibnamefont{Wells}} \bibnamefont{et~al.}
  (\bibinfo{collaboration}{SAMPLE Collaboration}), \bibinfo{journal}{Phys.
  Rev.} \textbf{\bibinfo{volume}{C63}}, \bibinfo{pages}{064001}
  (\bibinfo{year}{2001}); \bibinfo{author}{\bibfnamefont{E.~J.} \bibnamefont{Beise}},
  \bibinfo{author}{\bibfnamefont{M.~L.} \bibnamefont{Pitt}}, \bibnamefont{and}
  \bibinfo{author}{\bibfnamefont{D.~T.} \bibnamefont{Spayde}},
  \bibinfo{journal}{Prog. Part. Nucl. Phys.} \textbf{\bibinfo{volume}{54}},
  \bibinfo{pages}{289} (\bibinfo{year}{2005}).

\bibitem[{\citenamefont{Maas et~al.}(2005)}]{Maas:2004pd}
\bibinfo{author}{\bibfnamefont{F.~E.} \bibnamefont{Maas}} \bibnamefont{et~al.}
  (\bibinfo{collaboration}{A4 Collaboration}), \bibinfo{journal}{Phys. Rev.
  Lett.} \textbf{\bibinfo{volume}{94}}, \bibinfo{pages}{082001}
  (\bibinfo{year}{2005}).

\bibitem[{\citenamefont{{L. Capozza}}()}]{Capozza}
\bibinfo{author}{\bibnamefont{{L. Capozza}}} (\bibinfo{collaboration}{A4
  Collaboration}), \bibinfo{howpublished}{{PAVI06 proceedings, pending}}; 
\bibinfo{author}{\bibnamefont{{L. Kaufman}}} (\bibinfo{collaboration}{HAPPEX
  Collaboration}), \bibinfo{howpublished}{{PAVI06 proceedings, pending}}.



\bibitem[{\citenamefont{Gorchtein}(2006{\natexlab{b}})}]{Gorchtein:2005za}
\bibinfo{author}{\bibfnamefont{M.}~\bibnamefont{Gorchtein}},
  \bibinfo{journal}{Phys. Rev.} \textbf{\bibinfo{volume}{C73}},
  \bibinfo{pages}{035213} (\bibinfo{year}{2006}{\natexlab{b}}).

\bibitem[{\citenamefont{Armstrong et~al.}(2005)}]{Armstrong:2005hs}
\bibinfo{author}{\bibfnamefont{D.~S.} \bibnamefont{Armstrong}}
  \bibnamefont{et~al.} (\bibinfo{collaboration}{G0 Collaboration}),
  \bibinfo{journal}{Phys. Rev. Lett.} \textbf{\bibinfo{volume}{95}},
  \bibinfo{pages}{092001} (\bibinfo{year}{2005}).

\bibitem[{\citenamefont{Armstrong et~al.}()}]{G0NIM}
\bibinfo{author}{\bibfnamefont{D.~S.} \bibnamefont{Armstrong}}
  \bibnamefont{et~al.} (\bibinfo{collaboration}{G0 Collaboration}),
  \bibinfo{journal}{(in preparation)}.

\bibitem[{\citenamefont{Covrig et~al.}(2005)}]{Covrig:2005vx}
\bibinfo{author}{\bibfnamefont{S.~D.} \bibnamefont{Covrig}}
  \bibnamefont{et~al.}, \bibinfo{journal}{Nucl. Instrum. Meth.}
  \textbf{\bibinfo{volume}{A551}}, \bibinfo{pages}{218} (\bibinfo{year}{2005}).

\bibitem[{\citenamefont{Marchand et~al.}()}]{G0_electronics_NIM}
\bibinfo{author}{\bibfnamefont{D.}~\bibnamefont{Marchand}}
  \bibnamefont{et~al.}, \bibinfo{journal}{nucl-ex/0703026 [Submitted to Nucl. Instrum.
  Meth. Phys. Res.]}.

\bibitem[{\citenamefont{Sinclair et~al.}(2007)}]{Sinclair:2007ez}
\bibinfo{author}{\bibfnamefont{C.~K.} \bibnamefont{Sinclair}}
  \bibnamefont{et~al.}, \bibinfo{journal}{Phys. Rev. ST Accel. Beams}
  \textbf{\bibinfo{volume}{10}}, \bibinfo{pages}{023501}
  (\bibinfo{year}{2007}).


\bibitem[{\citenamefont{Hauger et~al.}(2001)}]{Hauger:1999iv}
\bibinfo{author}{\bibfnamefont{M.}~\bibnamefont{Hauger}} \bibnamefont{et~al.},
  \bibinfo{journal}{Nucl. Instrum. Meth.} \textbf{\bibinfo{volume}{A462}},
  \bibinfo{pages}{382} (\bibinfo{year}{2001}).


\bibitem[{afa()}]{afanasev_rad}
\bibinfo{title}{The correction is being investigated by A.~V.~Afanasev and N.~P.~Merenkov (private communication)}.


\end{thebibliography}
\end{document}